\documentclass[12pt]{iopart}
\usepackage[utf8]{inputenc}

\expandafter\let\csname equation*\endcsname=\relax 
\expandafter\let\csname endequation*\endcsname=\relax 
\usepackage{amsmath,amssymb,amsthm} 
\usepackage{braket}
\usepackage{graphicx}
\usepackage{hyperref}
\usepackage{xcolor}
\DeclareUnicodeCharacter{2264}{\ensuremath{\leq}}
\DeclareUnicodeCharacter{2009}{\,}

\usepackage[
    backend=biber,
    style=numeric-comp, 
    sorting=none,       
    maxcitenames=2,
    maxbibnames=99,
    giveninits=false
]{biblatex}
\begin{document}

\title{Visualising Quantum Entanglement Using Interactive Electronic Quantum Dice}

\author{B. Folkers$^1$, A. van Rossum$^1$, A Brinkman$^1$ and 
H. K. E. Stadermann$^{1,2}$}

\address{$^1$ MESA+ Institute for Nanotechnology, University of Twente, 7500 AE, Enschede, The Netherlands}
\address{$^2$ ELAN Institute for Teacher Training, University of Twente, 7500 AE, Enschede, The Netherlands}

\ead{b.folkers@utwente.nl}
\vspace{10pt}
\begin{indented}
\item[]October 2025
\end{indented}

\begin{abstract} Quantum entanglement remains a challenging concept to teach and visualise due to its microscopic and non-classical nature. We present innovative educational demonstration material consisting of electronic dice that simulate the properties of quantum entanglement through haptic interaction. The system uses displays, orientation sensors, and wireless communication to visualise key quantum mechanical principles such as superposition, measurement, and entanglement correlations. This analogy enables students to experience quantum phenomena through familiar objects, making abstract concepts more tangible. The Dice support various educational scenarios, from basic entanglement demonstrations to more complex quantum key distribution experiments, and can be adapted for different educational levels from secondary school to undergraduate physics courses. Initial implementations demonstrate that the interactive nature of the Quantum Dice can help users develop an intuitive understanding of quantum mechanical principles. The low-cost, open source, and robust design makes Quantum Dice accessible to a wider range of educational institutions.
\end{abstract}

\vspace{2pc}

\submitto{\PED}

\maketitle


\section{Introduction} \label{introduction}
Teaching Quantum Physics (QP) is challenging due to the abstract and mathematical nature of its models and the inability to interact directly with quantum phenomena \parencite{anupam_design_2020, corsiglia_intuition_2023, dreyfus_splits_2019}. This is particularly evident for Quantum Entanglement (QE), as it involves microscopic correlations that contradict classical expectations. Students struggle with QE as it builds on other abstract QP concepts such as superposition, quantum states, and quantum measurement. In addition, teachers find it challenging to access or develop teaching resources and learning activities on QP topics \parencite{satanassi_quantum_2021, pallotta_bringing_2022}. Especially QE is difficult to visualise using macroscopic objects with classical correlations \parencite{kaur_teaching_2017, aehle_approach_2022}. 

In recent years, different initiatives have been launched to address the challenge of visualising QP and QE concepts through the use of demonstration materials (DM) \parencite{seskir_quantum_2022, krijtenburg-lewerissa_insights_2017, singh_review_2015}. Current approaches include computer simulations \parencite{kohnle_investigating_2014, migdal_visualizing_2022}, analogies \parencite{andreotti_teaching_2022}, single-photon optical experiments \parencite{beck_quantum_2014, borish_seeing_2023},  virtual laboratories \parencite{migdal_visualizing_2022}, games \parencite{lopez-incera_entangle_2019}, and interactive classroom simulations \parencite{la_cour_virtual_2022}. These approaches offer complementary benefits: digital materials excel in visualisation and multiple representations, while physical materials create tactile engagement and authentic experimental contexts.

In this paper, we describe a new haptic electronic analogy, Quantum Dice, designed to simulate basic QP principles. The Dice enable students to explore and model QP concepts such as quantum states, superposition, quantum measurement, and entanglement through recognisable objects. The system supports various educational scenarios, from basic entanglement demonstrations to more sophisticated quantum key distribution experiments, and can be adapted for different educational levels from the general public to secondary school and undergraduate physics courses. In this article, we discuss the educational rationale, design principles, technical implementation, and educational applications of the Quantum Dice system. 

\section{Basic Operation of Quantum Dice} \label{working_principle}
A Quantum Die is made from a 3D printed frame equipped with displays on the faces of the die. The die represents a six-state quantum object. Before rolling, each face displays all possible outcomes simultaneously, resembling a superposition state. After the die is rolled, the top face displays a value between 1 and 6 with equal probability. Rolling the die and reading the top face represents a measurement, comparable to a normal die. The die can land in one of the three colours: red, yellow, and blue, each representing a different measurement basis.

\begin{figure*} [h]
    \centering
    \includegraphics[width=1\linewidth]{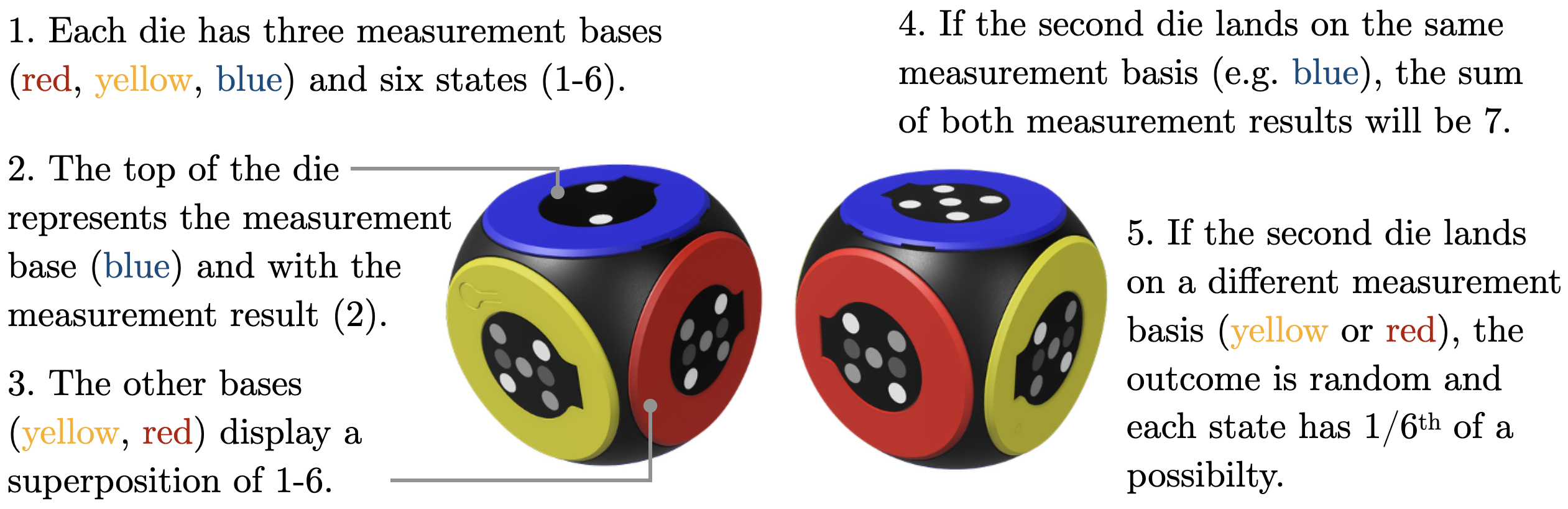}
    \caption{Operation of the entanglement mode of the Quantum Dice.}
    \label{working principle dice entangled}
\end{figure*}

When two Dice are brought close together, the displays change colour and represent an entangled state. If both Dice are rolled with the same colour on top, the sum of both outcomes will always be 7, although the individual outcomes will remain unpredictable (see Fig. \ref{working principle dice entangled}). In other words, the measurement statistics of one die depend on the measurement basis and measurement outcome of the other die. In a classical system, the outcomes of both Dice would be completely independent and the choice of colour would have no impact on the outcomes. See Section \ref{working_mech} for a more detailed description. In addition, an instructional video of the Dice is available on our website through the supplementary materials.

\section{Educational Rationale}\label{rationale}
Previous approaches to demonstrating quantum concepts often rely on elaborate optical setups or entirely virtual simulations. The Quantum Dice offers a tangible alternative that combines physical interaction with simulations of quantum physics concepts. Microscopic QP effects are typically difficult to observe directly, leading to instruction that relies on direct explanation rather than exploratory learning experiences.

The Dice provide opportunities for Model Based Learning, enabling users to construct, test, and refine mental models of QP phenomena \cite{buckleyModelBasedTeachingLearning2004}. Various demonstration materials in QP education align with these Model Based Learning principles \parencite{mckagan_developing_2008, sales_activities_2008, buongiorno_one_2018, anupam_particle_2018, ubbenExploringRelationshipStudents2023, xenakisQuantumSeriousGames2023}. Through interaction with the dice, students can develop explanations through analogical reasoning, where familiar and visible base domains support understanding of abstract concepts \parencite{aubussonMetaphorAnalogyScience2006a, gentnerStructuremappingTheoreticalFramework1983}. This is particularly valuable in QP education, where analogies could help make unfamiliar concepts accessible \parencite{andreotti_teaching_2022, rodriguezRoleAnalogiesClassical2025}.

The haptic nature of the Quantum Dice aligns with embodied cognition principles, serving as material anchors that ground abstract concepts in physical experience \parencite{weisberg_embodied_2017, hutchinsMaterialAnchorsConceptual2005}. The Dice could support the reduction of cognitive load through multimodal sensory engagement using visual, tactile, and spatial channels, and enable cognitive offloading by serving as objects that externally represent parts of the reasoning process \parencite{kirshDistinguishingEpistemicPragmatic1994, castro-alonsoResearchAvenuesSupporting2024}. Such embodied cognition approaches have been particularly valuable in QP education, while also increasing student motivation \parencite{schiber_student_2013, lopez-incera_entangle_2019, lopez-incera_encrypt_2020, marckwordt_entanglement_2021, zable_investigating_2020}.
Finally, we aligned our design with the spin-first approach to teaching QP. This approach uses simple, small quantum systems, such as spin with two levels, to help students focus on conceptual understanding while minimising complex mathematics early in the learning process \parencite{sadaghiani_spin_2015, michelini2000proposal, manogue2012representations, dur_visualization_2014, dur_qubit_2016}.

\section{Educational Design Principles} \label{requirement}
Following the educational rationale outlined in Section \ref{rationale}, we outline a set of educational principles that guided the development of the Quantum Dice. These principles appear across recent DM publications, and this previous work has equally inspired our design approach. Therefore, we cite relevant publications for each principle to show how these principles appear in different approaches and to acknowledge the research foundation underlying our design.

\begin{enumerate}
    \item \textbf{Empirical} \parencite{kohnle_enhancing_2015, marshman_investigating_2017, vlachopoulos_effect_2017}: The Dice are designed to allow students to empirically experience how entangled objects can be correlated. This approach may allow learners to model the observed correlations by enabling them to formulate hypotheses, perform tests, and refine their understanding based on the observed outcomes.
    
    \item \textbf{Haptic} \parencite{weisberg_embodied_2017, aehle_approach_2022, kaur_teaching_2017, liao_interactive_2022}: The Dice are designed as physical haptic materials, making abstract concepts more tangible and supporting learning through multisensory interaction. The use of haptic materials may facilitate teaching approaches that involve collective student interaction, where Dice act as material anchors.
        
    \item \textbf{Accessible} \parencite{haverkamp_simple_2022, khandelwal_cost-effective_2021, kaur_teaching_2017}:
    The Dice are designed to be used in a typical classroom setting and should not require prior knowledge of the technical instruments. The Dice design also prioritises cost-effectiveness.
    
    \item \textbf{Intuitive} \parencite{aehle_approach_2022, gordon_quantum_2012}: The Dice design requires no prior knowledge of technical instruments and builds upon familiar contexts to facilitate comprehension. Through interaction with the Dice, users without physics backgrounds should intuitively experience key distinctions from familiar classical behaviour of objects.
\end{enumerate}

\subsection{Concepts of Quantum Entanglement}
We aimed to simulate foundational concepts of quantum entanglement through our developed demonstration material. Basic explanations of these concepts are available in the work of Pade and Nielsen \& Chuang \parencite{pade2014quantum, nielsen2010quantum}. 

\begin{enumerate}
    \item \textbf{Multiple quantum objects, one quantum state}: Multiple entangled quantum objects are described by a single quantum state, regardless of the spatial separation of the quantum objects.
    
    \item \textbf{Superposition}: Entangled objects exist in a superposition state. Furthermore, a single quantum state can be expressed as a superposition of orthogonal states in a certain measurement bases.
    
    \item \textbf{Measurement}: Measurement of one entangled quantum object collapses the superposition state, instantaneously determining the state of its entangled counterparts. The choice of measurement basis influences the correlation statistics between entangled systems.
    
    \item \textbf{Statistical Measurement Results}: Measurements in entangled systems produce probabilistic rather than deterministic results, appearing as statistical distributions. Measurement of object A enables probabilistic predictions about measurements of object B.

\end{enumerate}

\section{Technical Design and Implementation}
\subsection*{Technical Principles}

The Quantum Dice were designed with the following technical principles in mind:

\begin{enumerate}
\item \textbf{Mechanical Durability}: Designed to withstand repeated rolling and mechanical impacts.
\item \textbf{User Interface}: Minimalistic two-button interface with smooth, flicker-free display updates for clear visual feedback.

\item \textbf{Cost-Effectiveness}: Uses 3D printed components and standard electronics (approximately €200 per two-Dice set) to ensure accessibility and reproducibility.

\item \textbf{Reliability}: The system uses two custom PCBs. One board houses the microcontroller, displays, and sensors. The other handles the power supply. This design enhances system integration, minimises connection failures, and improves long-term reliability.

\item \textbf{Wireless Communication}: Processor enables Dice proximity detection and information sharing while maintaining battery efficiency.

\item \textbf{Open Source}: Complete design files and software available on GitLab for community use and development.

\end{enumerate}

\begin{figure} [!b]
\centering
\includegraphics[width=0.7\linewidth]{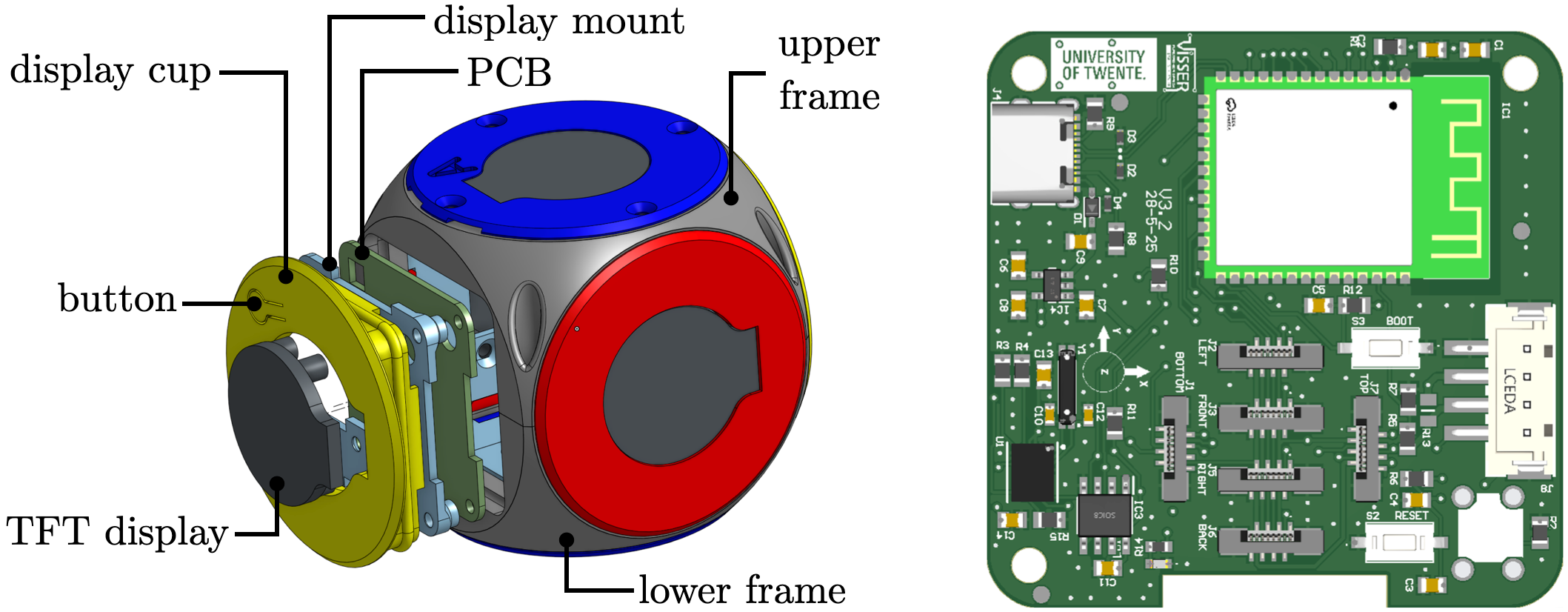}
\caption{Expanded view and one of the PCB's of the Quantum Dice. The frame is shaped like a dice with coloured cups in which the TFT display are mounted. The PCB's are mounted on the back side of the cups.}
\label{fig:dice exploded view}
\end{figure}

\subsection*{Construction of the Quantum Dice}
A Quantum Die consists of 3D printed components and electronic parts. The spherical frame is made from a flexible material with six flattened surfaces, each containing a circular LCD display in coloured cups (red, yellow, blue) as shown in Fig. \ref{fig:dice exploded view}. A die measures 76 x 76 x 76 mm³.

Each die contains a microcontroller, six displays, motion sensors, and a cryptographic chip for generating random numbers. The system operates on a rechargeable battery and includes wireless communication to detect a nearby die and share information. Two custom circuit boards house the electronics. Complete technical specifications, 3D printing files, circuit board designs, and software code are provided in the supplementary materials.
\\

\section{Operation and Applications} \label{working_mech}
This section details the operation of the Quantum Dice, from the basic mechanism of a single die to the simulation of entanglement with a pair of dice. In addition, it explains their practical application in teaching quantum key distribution protocols.

\subsection{Single Die Operation} \label{single_die_operation}
\begin{figure} [h]
    \centering
    \includegraphics[width=0.7\linewidth]{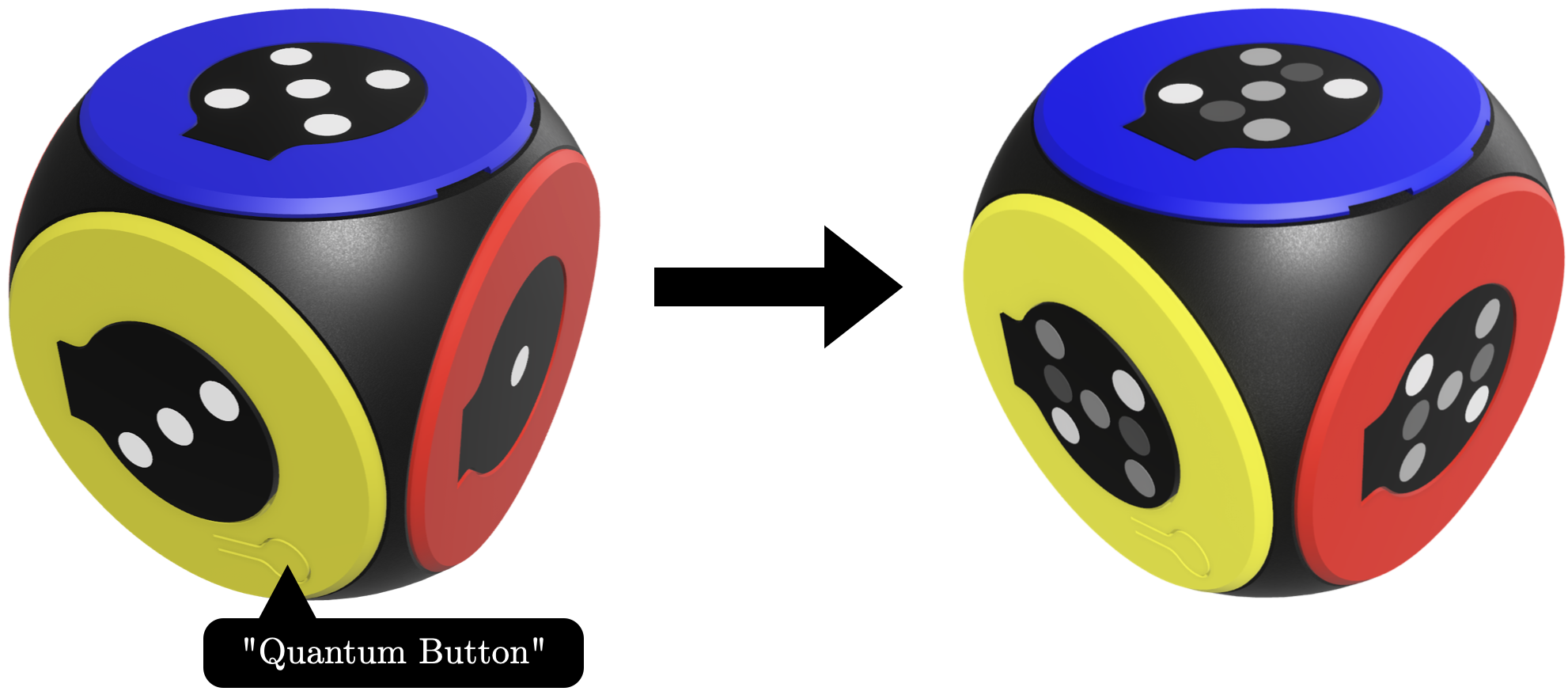}
    \caption{Activating the visual representation of the superposition state, with overlapping numbers indicating all possible outcomes.}
    \label{single_die_sup}
\end{figure}

A single die can simulate quantum superposition and measurement. When a die is turned on, the displays present a fixed number on each face of the dice, comparable to normal dice. Pressing the `Quantum Mode' button activates the superposition state, where each face visually represents overlapping semitransparent numbers. The different measurement bases are represented by red, blue and yellow colours (see Fig. \ref{single_die_sup}). 

We can represent this initial state mathematically:
\begin{equation}
    \ket{\psi} = \frac{1}{\sqrt{6}}(\ket{1} + \ket{2}+ \ket{3}+ \ket{4}+ \ket{5}+ \ket{6})
\end{equation}

\begin{figure} [!h]
    \centering
    \includegraphics[width=0.9\linewidth]{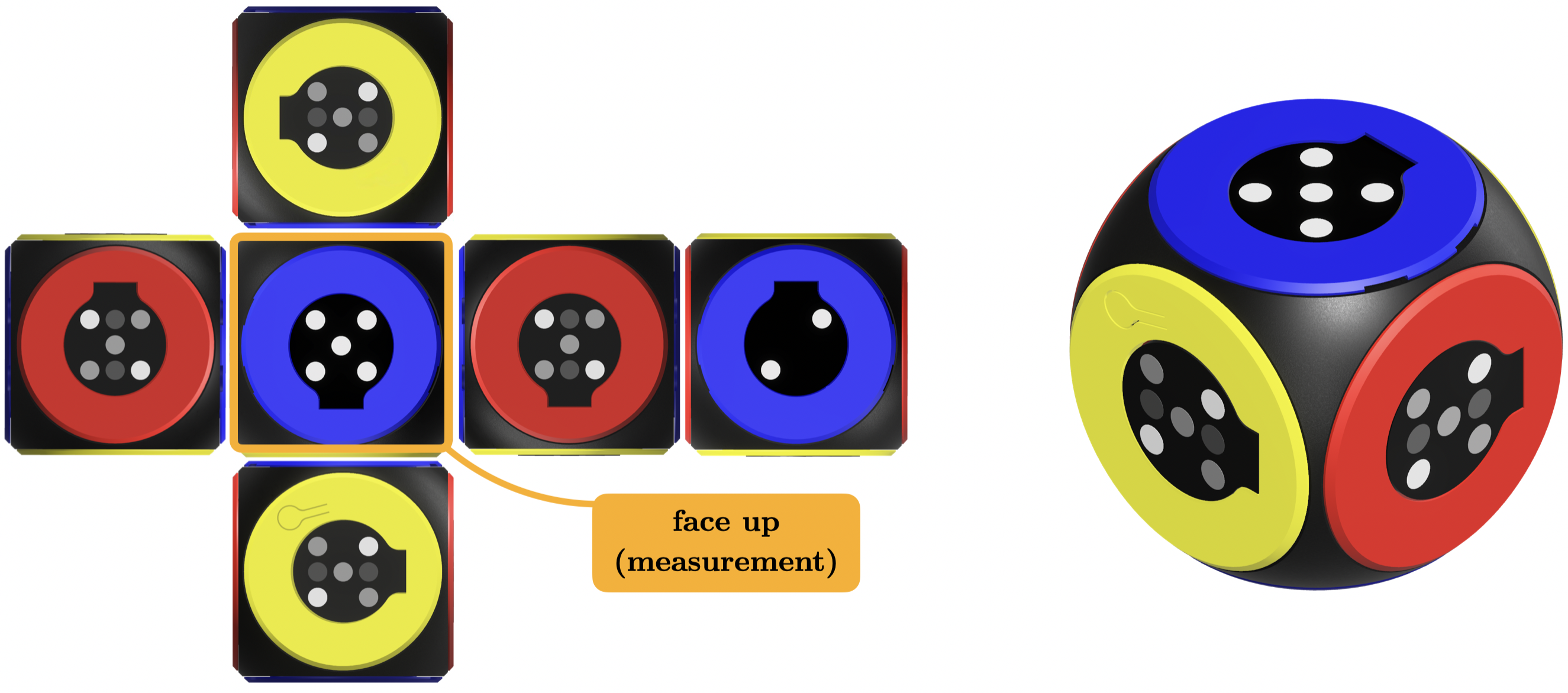}
    \caption{State 5  has been measured in the blue basis. A superposition remains in basis red and yellow.}
    \label{single_die_meas_1}
\end{figure}

Rolling the die simulates a measurement. The face of the die facing upward represents the measurement outcome in its basis. For example, if blue faces up, we consider the measurement to be performed in the blue basis (cf. Fig. \ref{single_die_meas_1}). 

\begin{figure} [b]
    \centering
    \includegraphics[width=0.9\linewidth]{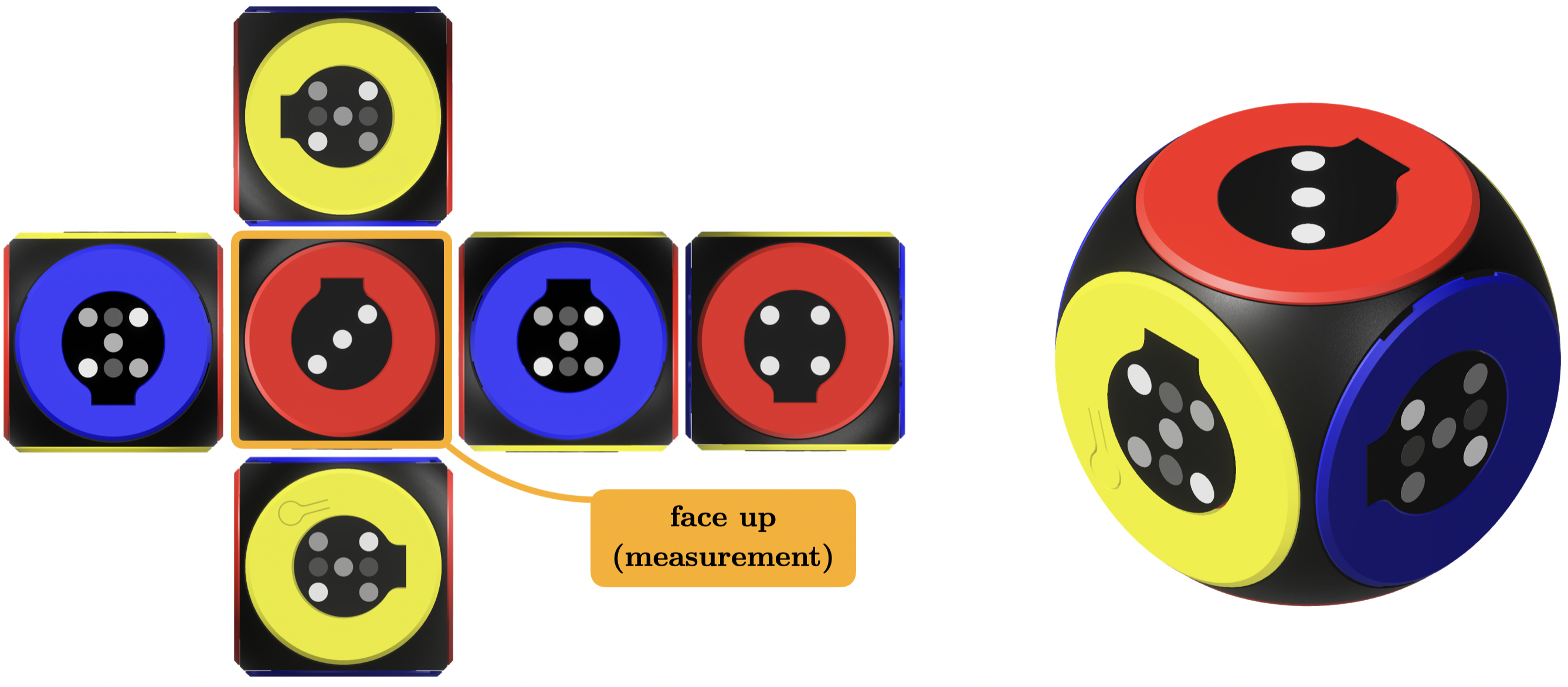}
    \caption{When the same Quantum Die is rolled and a different colour is facing up, the measurement result will again be unpredictable.}
    \label{single_die_meas_2}
\end{figure}

A single number from 1 to 6 is then displayed, representing the measurement outcome in that basis, with each outcome having a programmed probability of 1/6. Meanwhile, the other faces (red and yellow) continue to represent a superposition of all possible outcomes. The state of the die after this measurement can be described mathematically as follows:

\begin{equation}
    \ket{\psi} = \ket{5}_B  \,\,\,\,\,\,\text{and} \,\,\,\,\, \ket{\psi} = \frac{1}{\sqrt{6}}(\ket{1}_{R/Y} + \ket{2}_{R/Y} + \ket{3}_{R/Y}+ \ket{4}_{R/Y}+ \ket{5}_{R/Y}+ \ket{6}_{R/Y}) 
\end{equation}

When the die is rolled again, there are essentially two options. If the die lands on the same colour (blue in this case), the same outcome will be displayed, representing that measuring a quantum state in the same basis will yield the same result. If the die lands in another colour, such as red or yellow, the outcome will be unpredictable and determined by the coefficients of the superposition state (cf. Fig. \ref{single_die_meas_2}). 

This behaviour is analogous to measuring a quantum state in different bases, for example, with the polarisation of photons or electron spin \parencite{toninelli_concepts_2019, justice_improving_2019}.

\subsection{Simulating Quantum Entanglement with two Quantum Dice}
\begin{figure} [!h]
    \centering
    \includegraphics[width=0.5\linewidth]{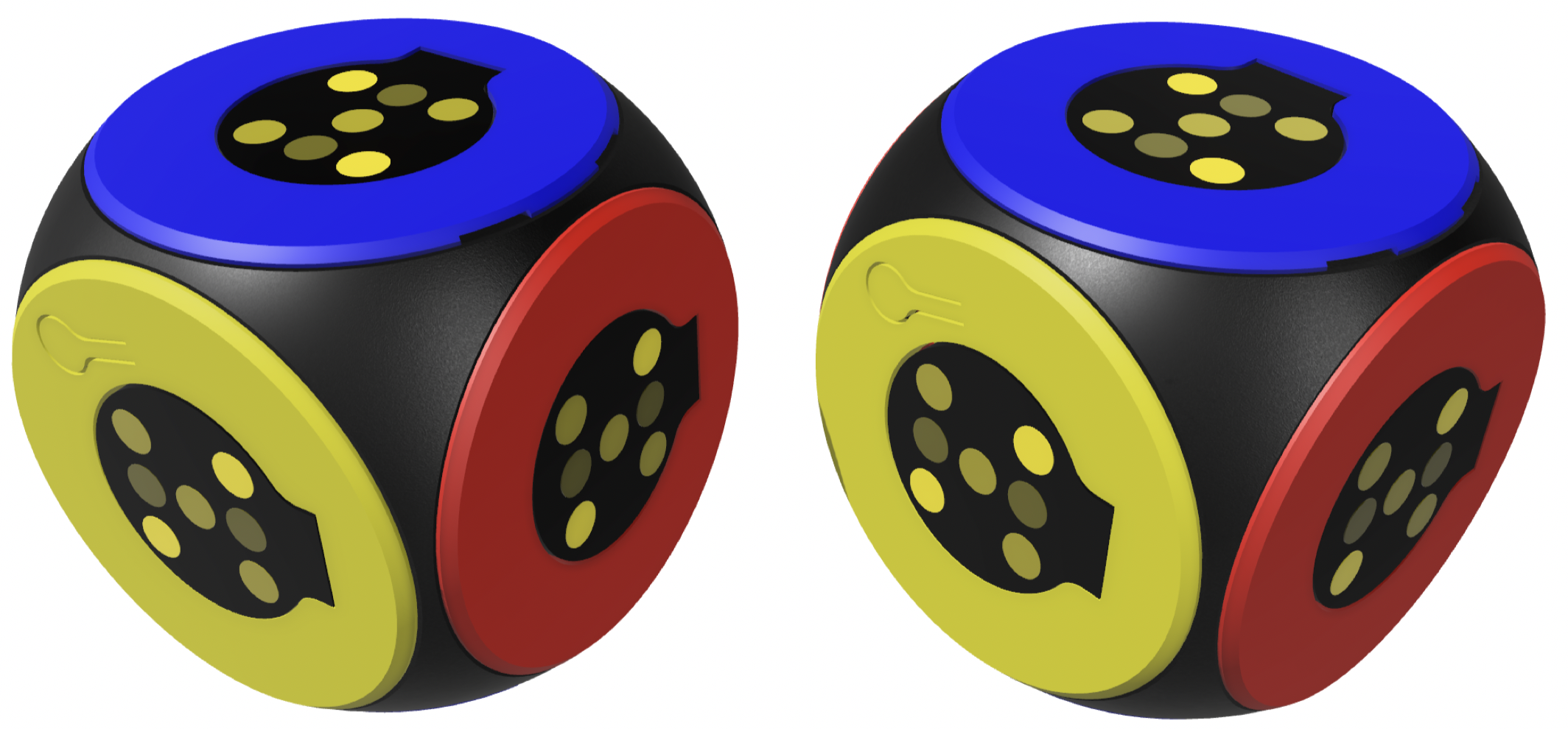}
    \caption{Yellow superposition symbols represent an entangled state when the Dice are brought within a few centimetres.}
    \label{two_dice_entangled_1}
\end{figure}

Bringing two Quantum Dice within a few centimetres of each other triggers the transition into an `entangled state.' This is indicated by the die-eyes of the superposition symbols turning yellow on both Dice (see Fig. \ref{two_dice_entangled_1}). The shared colour visualises the entangled state as a superposition state that is shared between the two dice. To explore what this entangled state implies for the behaviour of the dice, the Dice must be rolled again. Through repeated rolling of the entangled dice, the resulting correlations allow users to reason about the consequences of this shared quantum state.

In the 'entangled state', both Dice represent a single quantum system comprising two qubits, die A and die B. Mathematically, we can express the entangled state as follows:
\begin{equation}
    \ket{\psi}_{AB} = \frac{1}{\sqrt{6}}(\ket{1}_A\ket{6}_B + \ket{2}_A\ket{5}_B + \ket{3}_A\ket{4}_B+ \ket{4}_A\ket{3}_B+ \ket{5}_A\ket{2}_B+ \ket{6}_A\ket{1}_B)
    \label{entangled_equation}
\end{equation}

\begin{figure} [!b]
    \centering
    \includegraphics[width=0.5\linewidth]{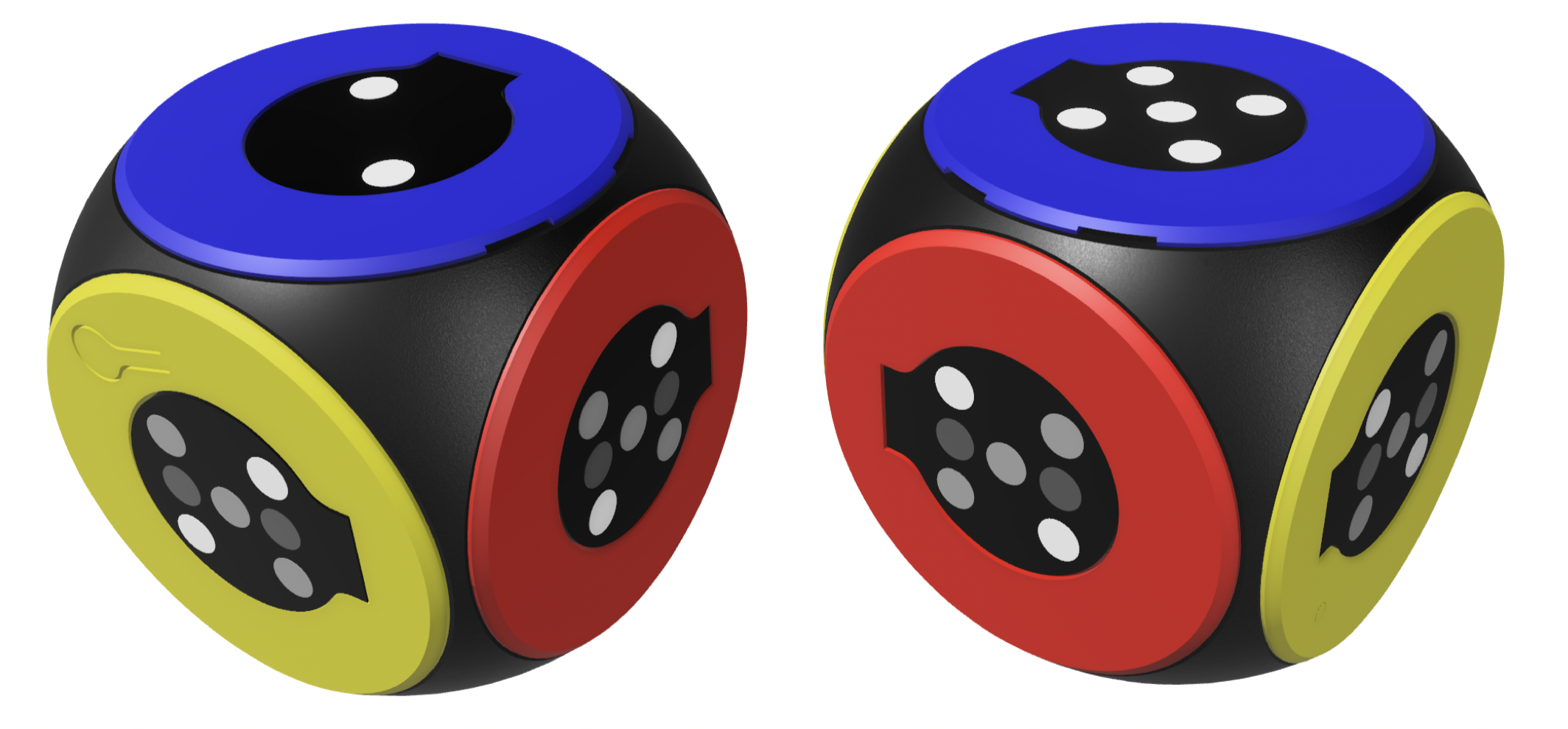}
    \caption{If entangled Dice are measured in the same basis colour, the sum of outcomes will be seven.}
    \label{two_dice_entangled_2}
\end{figure}

The Quantum Dice exhibit anti-correlation, although this can be programmed differently. This means that when both Dice are rolled in the same colour, the outcomes are never identical but always sum up to 7 (see \ref{two_dice_entangled_2}). This behaviour is analogous to opposite sides of a normal die, which also sum to 7. In this representation, the Dice behave as a single quantum system, where the measurement basis (colour) determines the correlation pattern. However, which individual Die will display which specific outcome remains uncertain until measurement. One only knows that their combined outcomes will sum to 7 when measured in the same colour.

\begin{figure} [!h]
    \centering
    \includegraphics[width=0.5\linewidth]{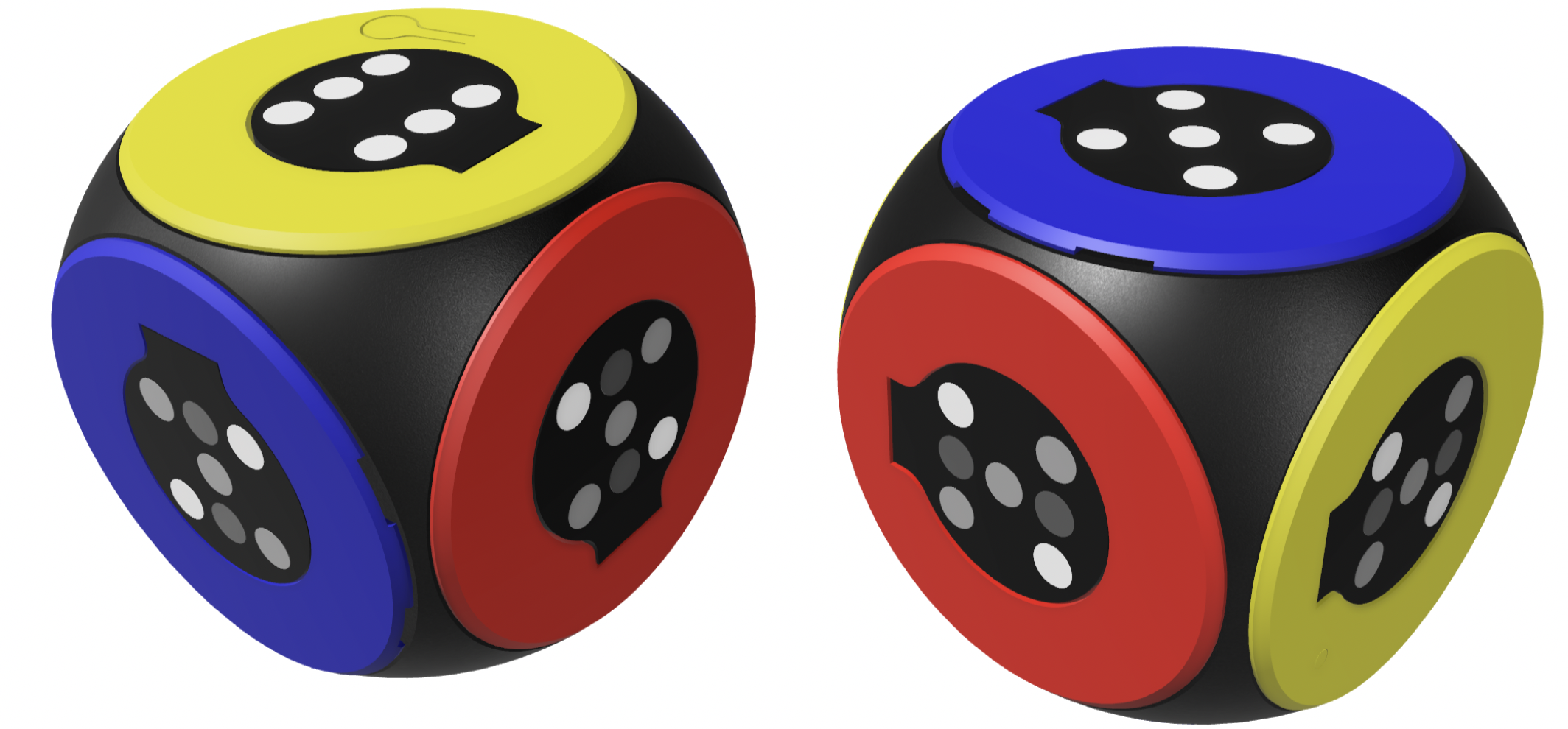}
    \caption{If entangled Dice are measured in a different basis colour with respect to each other, the outcome will be unpredictable.}
    \label{two_dice_entangled_3}
\end{figure}

If both Dice are rolled in different colours, the outcomes are uncorrelated. In this case, both Dice can display any value with equal probability (e.g., Fig. \ref{two_dice_entangled_3}). When the Dice are rolled while in the 'entangled state', the colour changes back to white, indicating that the Dice are no longer entangled (cf. Fig. \ref{two_dice_entangled_2} and \ref{two_dice_entangled_3}). The Dice must be brought into close proximity again to be prepared in an entangled state. This behaviour represents how entangled states collapse in authentic quantum systems upon measurement.

\subsection{Simulating Quantum Key Distribution with Quantum Dice}

\begin{figure} [!b]
\centering
\includegraphics[width=0.7\linewidth]{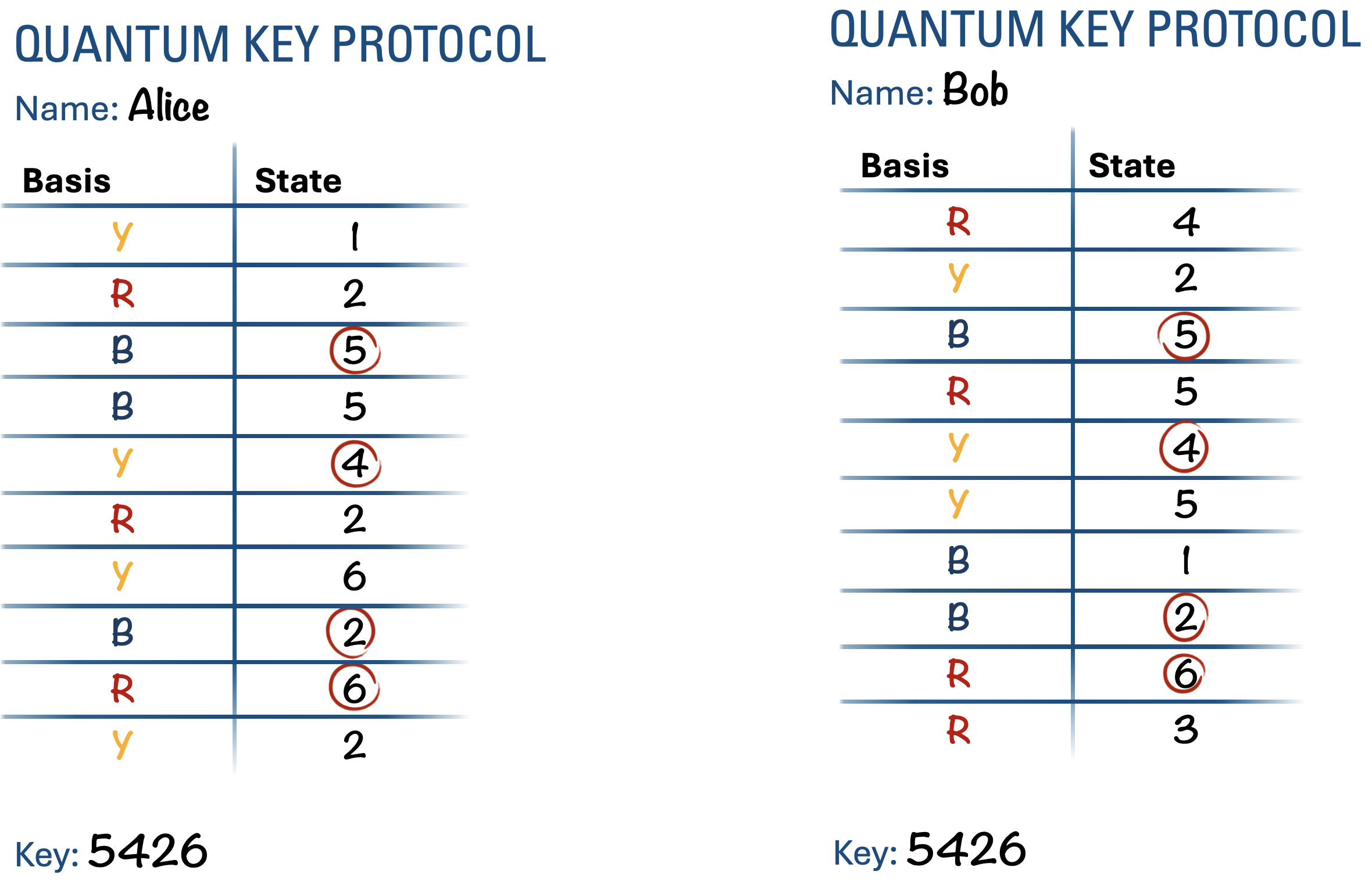}
\caption{Example of key formation using single Quantum Die measurements.}
\label{single-die-QKD-2}
\end{figure}

The Quantum Dice can demonstrate simplified quantum key distribution protocols. Using a single die (BB84 protocol \parencite{bennettQuantumCryptographyPublic2014}), users A and B take turns rolling the die and recording both the measurement basis colour and outcome (1-6). After each measurement by user B, the "Quantum Mode" button resets the die to superposition, simulating a new quantum state. This process repeats multiple times. When users share their measurement bases (not the outcomes), they discard the results of measurements performed in different bases (i.e., with different colours). Matching colours yield identical outcomes, as measuring a quantum state in the same basis produces consistent results. These matching outcomes form a shared cryptographic key without explicitly transmitting the key values (Fig. \ref{single-die-QKD-2}). In educational settings, a third person (Eve) can be introduced as an eavesdropper who performs measurements while the die travels from Alice to Bob, demonstrating how quantum measurement disturbs the state and reveals the presence of interception.

\begin{figure} [!h]
\centering
\includegraphics[width=0.7\linewidth]{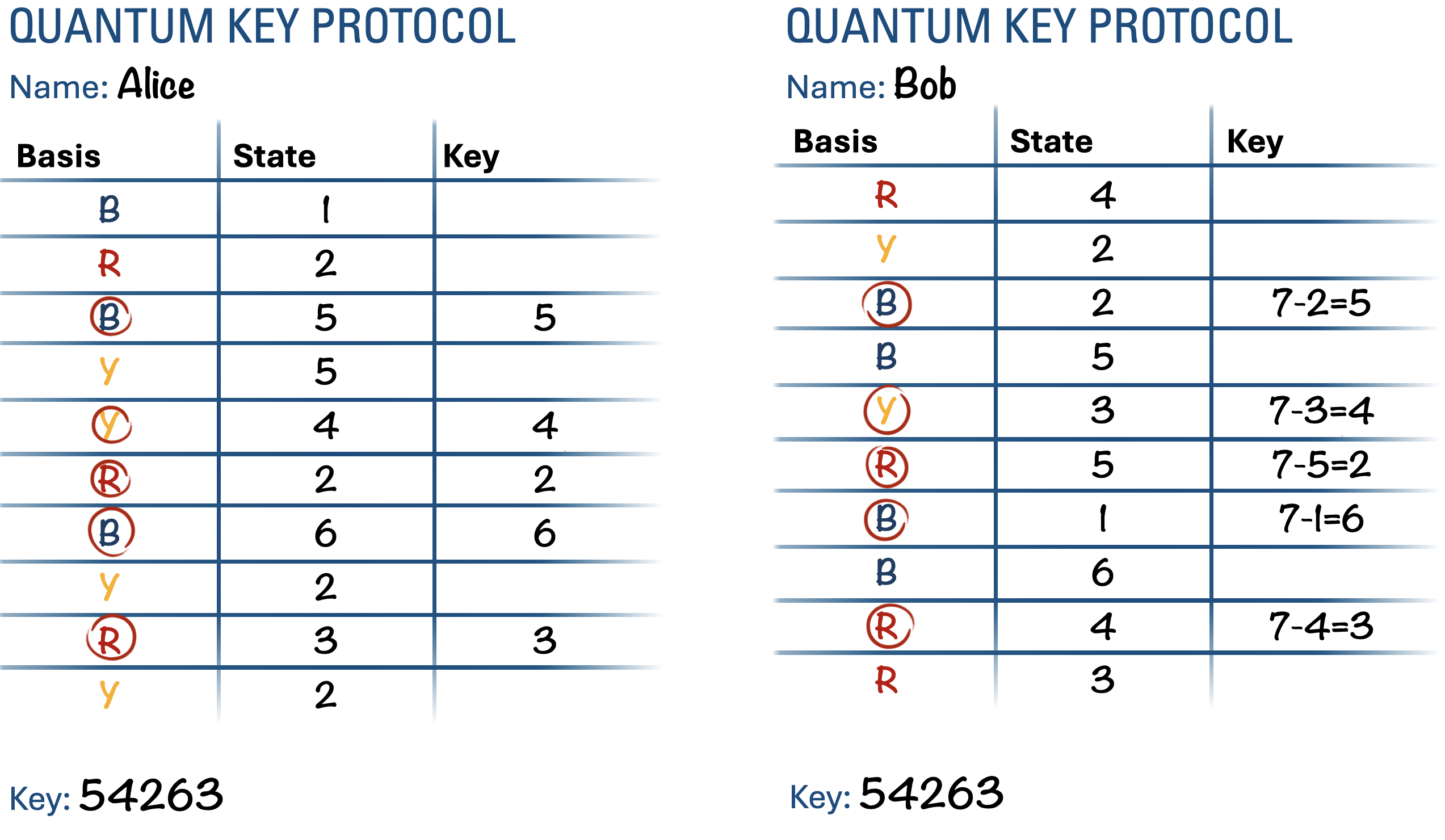}
\caption{Key formation using entangled Dice with re-entanglement between measurements.}
\label{QKD-1}
\end{figure}

This protocol can also be performed with two entangled Dice (E91 protocol \parencite{ekert1991quantum}). Users A and B each roll their own die simultaneously, recording colours and outcomes. After sharing measurement bases, they discard mismatched colours. For matching colours, the anticorrelation ensures outcomes sum to 7, allowing both users to derive the other's value and establish a shared key (Fig. \ref{QKD-1}). This eliminates the chance for an Eve to eavesdrop, since there is no physical exchange of Dice between Alice and Bob, demonstrating a fundamental security advantage of entanglement-based protocols. Re-entanglement is required between rounds to prepare new entangled states.Through this correlation, they can establish a cryptographic key without explicitly transmitting it.

\section{Implementation, Experiences and Limitations}
Throughout the development and testing phases, we used the Quantum Dice with a variety of audiences, including participants at Dutch academic physics conferences, undergraduate lectures for non-physicists, and secondary school students. Our experiences during these activities suggest that the Dice can support users in reasoning about the entangled correlations. In university level settings, students spontaneously compared the Dice to physical quantum objects such as photons and electrons. These moments often led to discussion and reflection on QP concepts, particularly the role of measurement basis and statistics in quantum entanglement.

During demonstrations at public events, we also observed interest from participants without a physics background. On several occasions, users remarked on the perceived simplicity of the concepts being represented, as reflected in comments such as: \textit{``But this isn’t such a complex concept, right?’’} This reaction challenges the commonly held view of quantum entanglement as inherently difficult to explain. Consistent with this, users quickly engaged with the Dice and discussed the observed correlations. This engagement suggests that the Dice could provide a low-threshold entry point for discussing the non-classical effects represented by the system. Structured follow-up research could further substantiate these observations.

The examples above demonstrate that the Quantum Dice can be implemented across a broad range of educational contexts. We also developed a simplified colourless version in which the outcomes always sum to seven when the Dice are brought into proximity, providing accessible entry points for different age groups. For upper-level secondary students, all quantum processes described in this article can, in principle, be implemented. At university level, the Dice additionally support more formal treatments using conventional state notation and mathematical representations. Because the system is fully programmable, the relative orientations of the colour measurement bases can be modified. Different basis choices then lead to distinct measurement statistics, providing opportunities to explore the role of basis orientation.\\

An effective implementation of Quantum Dice as an educational analogy requires careful pedagogical scaffolding. As with analogies in general, it is important to explicitly communicate the mapping between the analogy and the target quantum system \cite{harrison2006teaching}. In addition, the limitations of the analogy should be explicitly discussed. For example, the Quantum Dice represent only maximally entangled states, rely on classical communication between the dice, and treat the measurement basis as an intrinsic property of the system, all of which differ from authentic quantum systems. To support educators, we have included a first structured overview of the explicit mappings, limitations, and potential undesired views in the Supplementary Information, which can serve as a starting point for reflection and further educational design research.

\section{Conclusion and Future Work}
The promising functionality demonstrated by the current prototype suggests the potential for further development and refinement. In the next year, our objective is to develop detailed practical instructions and teaching guides to further explore the applicability of the Quantum Dice. Future development may explore expanded quantum education applications, such as quantum teleportation protocols with a third die, a simplified CHSH experiment, and qubit representations.

We intend to investigate the educational impact of the Quantum Dice in a follow-up study. Specifically, investigating how learners explicitly map the mechanisms of the Dice onto authentic quantum systems and how they perceive and articulate these mechanisms would be valuable. Exploring how teachers might integrate the Dice into their lessons and how the Dice influences users' interpretation of QP phenomena could also provide important insights.

The source files for this project are open source. We encourage any further improvements, extensions, and discussion on this project. With the increasing demands in quantum (technology) education, this approach could aid the development of teaching materials for QP. Moreover, we believe that an open source approach is instrumental in accelerating innovation in QP education across schools, universities, and the general public.

\section*{Supplementary information}
Supplementary Information accompanying this article is available via the journal website. In addition, construction files and build instructions for the Quantum Dice are publicly available through a GitHub repository, see
\url{https://github.com/qlab-utwente/Quantum-Dice-by-UTwente}. An instructional video explaining the operation of the Quantum Dice can be accessed on the project website:
\url{https://ut.onl/quantumdice}.

\section*{Declarations}
\begin{itemize}
  \item \textbf{Funding} This research project is funded by the Quantum Delta NL Growth Fund.  
  \item \textbf{Conflict of interest/Competing interests} The authors declare that they have no competing interests and that they have no relevant financial or non-financial interests to disclose.
  \item \textbf{Materials availability} Designs, component-list and codes are available through the UTwente GitLab \url{hhttps://github.com/qlab-utwente/Quantum-Dice-by-UTwente}, or visit our website \url{https://ut.onl/quantumdice}.
  \item \textbf{Author contribution} B.F, developed the idea and working principle of the Dice and worked out the pedagogical considerations. A.v.R. designed the structure of the Dice and worked out the technical implementation. A.v.R. was responsible for the code and building of the dice. B.F. wrote the first draft of the manuscript. A.v.R, K.S. and A.B. reviewed the manuscript. All authors have read and agreed to the submitted version of the manuscript.
\end{itemize}

\newpage

\printbibliography

@book{pade2014quantum,
 author = {Pade, Jochen},
 publisher = {Springer},
 title = {Quantum mechanics for pedestrians 1: Fundamentals},
 year = {2014}
}

@book{nielsen2010quantum,
 author = {Nielsen, Michael A and Chuang, Isaac L},
 publisher = {Cambridge university press},
 title = {Quantum computation and quantum information},
 year = {2010}
}

@inproceedings{manogue2012representations,
 author = {Manogue, Corinne and Gire, Elizabeth and McIntyre, David and Tate, Janet},
 booktitle = {AIP Conference Proceedings},
 number = {1},
 organization = {American Institute of Physics},
 pages = {55--58},
 title = {Representations for a spins-first approach to quantum mechanics},
 volume = {1413},
 year = {2012},
 doi = {10.1063/1.3679992},
}

@article{michelini2000proposal,
 author = {Michelini, Marisa and Ragazzon, Renzo and Santi, Lorenzo and Stefanel, Alberto},
 journal = {Physics Education},
 number = {6},
 pages = {406},
 publisher = {IOP Publishing},
 title = {Proposal for quantum physics in secondary school},
 volume = {35},
 year = {2000},
 doi = {10.1088/0031-9120/35/6/305},
}

@article{krijtenburg-lewerissa_insights_2017,
 author = {Krijtenburg-Lewerissa, K. and Pol, H. J. and Brinkman, A. and van Joolingen, W. R.},
 doi = {10.1103/PhysRevPhysEducRes.13.010109},
 issn = {2469-9896},
 journal = {Physical Review Physics Education Research},
 number = {1},
 pages = {010109},
 title = {Insights into teaching quantum mechanics in secondary and lower undergraduate education},
 volume = {13},
 year = {2017}
}

@article{singh_review_2015,
 author = {Singh, Chandralekha and Marshman, Emily},
 doi = {10.1103/PhysRevSTPER.11.020117},
 issn = {1554-9178},
 journal = {Physical Review Special Topics - Physics Education Research},
 number = {2},
 pages = {020117},
 title = {Review of student difficulties in upper-level quantum mechanics},
 volume = {11},
 year = {2015}
}

@article{dur_qubit_2016,
 author = {Dür, Wolfgang and Heusler, Stefan},
 doi = {10.1119/1.4942137},
 issn = {0031-921X},
 journal = {The Physics Teacher},
 number = {3},
 pages = {156--159},
 shorttitle = {The {Qubit} as {Key} to {Quantum} {Physics} {Part} {II}},
 title = {The {Qubit} as {Key} to {Quantum} {Physics} {Part} {II}: {Physical} {Realizations} and {Applications}},
 volume = {54},
 year = {2016}
}

@article{khandelwal_cost-effective_2021,
 author = {Khandelwal, Aarushi and Tan, Jit Bin Joseph and Leong, Tze Kwang and Yang, Yarong and Venkatesan, T and Jani, Hariom},
 doi = {10.1088/1361-6552/abea49},
 issn = {0031-9120, 1361-6552},
 journal = {Physics Education},
 number = {3},
 pages = {033007},
 title = {A cost-effective quantum eraser demonstration},
 volume = {56},
 year = {2021}
}

@article{seskir_quantum_2022,
 author = {Seskir, Zeki C. and Migdał, Piotr and Weidner, Carrie and Anupam, Aditya and Case, Nicky and Davis, Noah and Decaroli, Chiara and Ercan, İlke and Foti, Caterina and Gora, Paweł and Jankiewicz, Klementyna and La Cour, Brian R. and Yago Malo, Jorge and Maniscalco, Sabrina and Naeemi, Azad and Nita, Laurentiu and Parvin, Nassim and Scafirimuto, Fabio and Sherson, Jacob F. and Surer, Elif and Wootton, James and Yeh, Lia and Zabello, Olga and Chiofalo, Marilù},
 doi = {10.1117/1.OE.61.8.081809},
 issn = {0091-3286},
 journal = {Optical Engineering},
 number = {08},
 title = {Quantum games and interactive tools for quantum technologies outreach and education},
 volume = {61},
 year = {2022}
}

@inproceedings{sadaghiani_spin_2015,
 address = {College Park, MD},
 author = {Sadaghiani, Homeyra R. and Munteanu, James},
 booktitle = {2015 {Physics} {Education} {Research} {Conference} {Proceedings}},
 doi = {10.1119/perc.2015.pr.067},
 pages = {287--290},
 publisher = {American Association of Physics Teachers},
 title = {Spin {First} instructional approach to teaching quantum mechanics in sophomore level modern physics courses},
 year = {2015}
}

@techreport{pallotta_bringing_2022,
 author = {Pallotta, Filippo},
 doi = {10.48550/arXiv.2206.15264},
 institution = {arXiv},
 note = {arXiv:2206.15264 [physics]
type: article},
 number = {arXiv:2206.15264},
 title = {Bringing the second quantum revolution into high school},
 year = {2022}
}

@article{marckwordt_entanglement_2021,
 author = {Marckwordt, Jasmine and Muller, Alexandria and Harlow, Danielle and Franklin, Diana and Landsberg, Randall H.},
 doi = {10.1119/5.0019871},
 issn = {0031-921X, 1943-4928},
 journal = {The Physics Teacher},
 number = {8},
 pages = {613--616},
 shorttitle = {Entanglement {Ball}},
 title = {Entanglement {Ball}: {Using} {Dodgeball} to {Introduce} {Quantum} {Entanglement}},
 volume = {59},
 year = {2021}
}

@article{justice_improving_2019,
 author = {Justice, Paul and Marshman, Emily and Singh, Chandralekha},
 doi = {10.1088/1361-6404/ab2135},
 issn = {0143-0807},
 journal = {European Journal of Physics},
 note = {Publisher: IOP Publishing},
 number = {5},
 pages = {055702},
 title = {Improving student understanding of quantum mechanics underlying the {Stern}–{Gerlach} experiment using a research-validated multiple-choice question sequence},
 volume = {40},
 year = {2019}
}

@article{kohnle_enhancing_2015,
 author = {Kohnle, Antje and Baily, Charles and Campbell, Anna and Korolkova, Natalia and Paetkau, Mark J.},
 doi = {10.1119/1.4913786},
 issn = {0002-9505, 1943-2909},
 journal = {American Journal of Physics},
 number = {6},
 pages = {560--566},
 title = {Enhancing student learning of two-level quantum systems with interactive simulations},
 volume = {83},
 year = {2015}
}

@article{dur_visualization_2014,
 author = {Dür, Wolfgang and Heusler, Stefan},
 doi = {10.1119/1.4897588},
 issn = {0031-921X},
 journal = {The Physics Teacher},
 note = {Publisher: American Association of Physics Teachers},
 number = {8},
 pages = {489--492},
 shorttitle = {Visualization of the {Invisible}},
 title = {Visualization of the {Invisible}: {The} {Qubit} as {Key} to {Quantum}                     {Physics}},
 volume = {52},
 year = {2014}
}

@article{mckagan_developing_2008,
 author = {McKagan, S. B. and Perkins, K. K. and Dubson, M. and Malley, C. and Reid, S. and LeMaster, R. and Wieman, C. E.},
 doi = {10.1119/1.2885199},
 issn = {0002-9505},
 journal = {American Journal of Physics},
 note = {Publisher: American Association of Physics Teachers},
 number = {4},
 pages = {406--417},
 title = {Developing and researching {PhET} simulations for teaching quantum mechanics},
 volume = {76},
 year = {2008}
}

@article{satanassi_quantum_2021,
 author = {Satanassi, Sara and Fantini, Paola and Spada, Roberta and Levrini, Olivia},
 doi = {10.1088/1742-6596/1929/1/012053},
 issn = {1742-6588, 1742-6596},
 journal = {Journal of Physics: Conference Series},
 number = {1},
 pages = {012053},
 shorttitle = {Quantum {Computing} for high school},
 title = {Quantum {Computing} for high school: an approach to interdisciplinary in {STEM} for teaching},
 volume = {1929},
 year = {2021}
}

@article{gordon_quantum_2012,
 author = {Gordon, Michal and Gordon, Goren},
 doi = {10.1088/0031-9120/47/3/346},
 issn = {0031-9120},
 journal = {Physics Education},
 number = {3},
 pages = {346},
 shorttitle = {Quantum computer games},
 title = {Quantum computer games: {Schrödinger} cat and hounds},
 volume = {47},
 year = {2012}
}

@article{andreotti_teaching_2022,
 author = {Andreotti, Erica and Frans, Renaat},
 doi = {10.1088/1361-6552/ac9ae3},
 issn = {0031-9120},
 journal = {Physics Education},
 note = {Publisher: IOP Publishing},
 number = {1},
 pages = {015008},
 title = {Teaching quantum physics in secondary schools using the analogy with the physics of musical instruments},
 volume = {58},
 year = {2022}
}

@article{aehle_approach_2022,
 author = {Aehle, Stefan and Scheiger, Philipp and Cartarius, Holger},
 copyright = {http://creativecommons.org/licenses/by/3.0/},
 doi = {10.3390/physics4040080},
 issn = {2624-8174},
 journal = {Physics},
 note = {Number: 4
Publisher: Multidisciplinary Digital Publishing Institute},
 number = {4},
 pages = {1241--1252},
 title = {An {Approach} to {Quantum} {Physics} {Teaching} through {Analog} {Experiments}},
 volume = {4},
 year = {2022}
}

@article{haverkamp_simple_2022,
 author = {Haverkamp, N. and Pusch, A. and Heusler, S. and Gregor, M.},
 doi = {10.1088/1361-6552/ac4106},
 journal = {Physics Education},
 number = {2},
 title = {A simple modular kit for various wave optic experiments using {3D} printed cubes for education},
 volume = {57},
 year = {2022}
}

@article{lopez-incera_encrypt_2020,
 author = {López-Incera, A. and Hartmann, A. and Dür, W.},
 doi = {10.1088/1361-6404/ab9a67},
 journal = {European Journal of Physics},
 number = {6},
 title = {Encrypt me! {A} game-based approach to {Bell} inequalities and quantum cryptography},
 volume = {41},
 year = {2020}
}

@article{lopez-incera_entangle_2019,
 author = {López-Incera, A. and Dür, W.},
 doi = {10.1119/1.5086275},
 journal = {American Journal of Physics},
 number = {2},
 pages = {95--101},
 title = {Entangle me! {A} game to demonstrate the principles of quantum mechanics},
 volume = {87},
 year = {2019}
}

@inproceedings{buongiorno_one_2018,
 author = {Buongiorno, D. and Michelini, M. and Santi, L. and Stefanel, A.},
 doi = {10.1088/1742-6596/1076/1/012011},
 note = {Issue: 1},
 title = {From one slit to diffraction grating: {Optical} physics lab by means of computer on-line sensors},
 volume = {1076},
 year = {2018}
}

@inproceedings{kohnle_investigating_2014,
 author = {Kohnle, A and Baily, C and Ruby, S},
 booktitle = {Proceedings of the Physics Education Research Conference},
 doi = {10.1119/perc.2014.pr.031},
 pages = {139--142},
 title = {Investigating the {Influence} of {Visualization} on {Student} {Understanding} of {Quantum} {Superposition}},
 year = {2014}
}

@article{sales_activities_2008,
 author = {Sales, G.L. and Vasconcelos, F.H.L. and Filho, J.A. de C. and Pequeno, M.C.},
 journal = {Revista Brasileira de Ensino de Fisica},
 number = {3},
 title = {Activities of exploratory modelling applied to the teaching of modern physics by using the learning object {The} {Quantum} {Duck}},
 volume = {30},
 year = {2008},
 doi = {10.1590/s1806-11172008000300017},
}

@article{marshman_investigating_2017,
 author = {Marshman, Emily and Singh, Chandralekha},
 journal = {Physical Review Physics Education Research},
 note = {Publisher: Physical Review Physics Education Research},
 number = {1},
 pages = {010117},
 title = {Investigating and {Improving} {Student} {Understanding} of {Quantum} {Mechanics} in the {Context} of {Single} {Photon} {Interference}},
 volume = {13},
 year = {2017},
 doi = {10.1103/PhysRevPhysEducRes.13.010117},
}

@article{anupam_design_2020,
 author = {Anupam, Aditya and Gupta, Ridhima and Gupta, Shubhangi and Li, Zhendong and Hong, Nora and Naeemi, Azad and Parvin, Nassim},
 journal = {International Journal of Designs for Learning},
 note = {Publisher: International Journal of Designs for Learning},
 number = {1},
 pages = {1--20},
 title = {Design {Challenges} for {Science} {Games}: {The} {Case} of a {Quantum} {Mechanics} {Game}},
 volume = {11},
 year = {2020},
 doi = {10.14434/ijdl.v11i1.24264},
}

@article{migdal_visualizing_2022,
 author = {Migdał, P. and Jankiewicz, K. and Grabarz, P. and Decaroli, C. and Cochin, P.},
 doi = {10.1117/1.OE.61.8.081808},
 journal = {Optical Engineering},
 number = {8},
 title = {Visualizing quantum mechanics in an interactive simulation-{Virtual} {Lab} by {Quantum} {Flytrap}},
 volume = {61},
 year = {2022}
}

@inproceedings{liao_interactive_2022,
 author = {Liao, Y.-P. and Cheng, Y.-L. and Zhang, Y.-T. and Wu, H.-X. and Lu, R.-C.},
 doi = {10.1109/QCE53715.2022.00097},
 pages = {718--723},
 title = {The interactive system of {Bloch} sphere for quantum computing education},
 year = {2022}
}

@inproceedings{la_cour_virtual_2022,
 author = {La Cour, B.R. and Maynard, M. and Shroff, P. and Ko, G. and Ellis, E.},
 booktitle = {Proceedings of the IEEE International Conference on Quantum Computing and Engineering (QCE)},
 doi = {10.1109/QCE53715.2022.00091},
 pages = {677--687},
 title = {The {Virtual} {Quantum} {Optics} {Laboratory}},
 year = {2022}
}

@inproceedings{zable_investigating_2020,
 author = {Zable, A. and Hollenberg, L. and Velloso, E. and Goncalves, J.},
 doi = {10.1145/3385956.3418957},
 title = {Investigating {Immersive} {Virtual} {Reality} as an {Educational} {Tool} for {Quantum} {Computing}},
 year = {2020}
}

@article{toninelli_concepts_2019,
 author = {Toninelli, E. and Ndagano, B. and Vallés, A. and Sephton, B. and Nape, I. and Ambrosio, A. and Capasso, F. and Padgett, M.J. and Forbes, A.},
 doi = {10.1364/AOP.11.000067},
 journal = {Advances in Optics and Photonics},
 number = {1},
 pages = {67--134},
 title = {Concepts in quantum state tomography and classical implementation with intense light: {A} tutorial},
 volume = {11},
 year = {2019}
}

@article{anupam_particle_2018,
 author = {Anupam, A. and Gupta, R. and Naeemi, A. and Jafarinaimi, N.},
 doi = {10.1109/TE.2017.2727442},
 journal = {IEEE Transactions on Education},
 number = {1},
 pages = {29--37},
 title = {Particle in a {Box}: {An} {Experiential} {Environment} for {Learning} {Introductory} {Quantum} {Mechanics}},
 volume = {61},
 year = {2018}
}

@article{kaur_teaching_2017,
 author = {Kaur, T. and Blair, D. and Moschilla, J. and Zadnik, M.},
 doi = {10.1088/1361-6552/aa83e1},
 journal = {Physics Education},
 number = {6},
 title = {Teaching {Einsteinian} physics at schools: {Part} 2, models and analogies for quantum physics},
 volume = {52},
 year = {2017}
}

@inproceedings{beck_quantum_2014,
 author = {Beck, M. and Dederick, E.},
 booktitle = {Proceedings of SPIE},
 doi = {10.1117/12.2070525},
 title = {Quantum optics laboratories for undergraduates},
 volume = {9289},
 year = {2014}
}

@inproceedings{xenakisQuantumSeriousGames2023,
 author = {Xenakis, A. and Avramouli, M. and Sabani, M. and Savvas, I. and Chaikalis, C. and Theodoropoulou, K.},
 doi = {10.1109/EDUCON54358.2023.10125266},
 eventtitle = {{{IEEE Global Engineering Education Conference}}, {{EDUCON}}},
 publisher = {IEEE Computer Society},
 title = {Quantum {{Serious Games}} to {{Boost Quantum Literacy}} within {{Computational Thinking}} 2.0 {{Framework}}},
 volume = {2023},
 year = {2023}
}

@article{rodriguezRoleAnalogiesClassical2025,
 author = {Rodriguez, Luiza Vilarta and Van Der Veen, Jan T. and De Jong, Ton},
 doi = {10.1103/PhysRevPhysEducRes.21.010108},
 issn = {2469-9896},
 journaltitle = {Physical Review Physics Education Research},
 langid = {english},
 number = {1},
 pages = {010108},
 shortjournal = {Phys. Rev. Phys. Educ. Res.},
 title = {Role of Analogies with Classical Physics in Introductory Quantum Physics Teaching},
 volume = {21},
 year = {2025}
}

@article{ubbenExploringRelationshipStudents2023,
 author = {Ubben, Malte and Bitzenbauer, Philipp},
 doi = {10.3389/frqst.2023.1207619},
 issn = {2813-2181},
 journaltitle = {Frontiers in Quantum Science and Technology},
 langid = {english},
 pages = {1207619},
 shortjournal = {Front. Quantum Sci. Technol.},
 title = {Exploring the Relationship between Students’ Conceptual Understanding and Model Thinking in Quantum Optics},
 volume = {2},
 year = {2023}
}

@inproceedings{schiber_student_2013,
 author = {Schiber, CC and Close, HG and Close, EW and Donnelly, D},
 doi = {10.1119/perc.2013.pr.069},
 pages = {325--328},
 title = {Student use of a material anchor for quantum wave functions},
 year = {2013}
}

@article{corsiglia_intuition_2023,
 author = {Corsiglia, Giaco and Pollock, Steven and Passante, Gina},
 doi = {10.1103/PhysRevPhysEducRes.19.010109},
 issn = {2469-9896},
 journal = {Physical Review Physics Education Research},
 number = {1},
 pages = {010109},
 shorttitle = {Intuition in quantum mechanics},
 title = {Intuition in quantum mechanics: {Student} perspectives and expectations},
 volume = {19},
 year = {2023}
}

@article{dreyfus_splits_2019,
 author = {Dreyfus, Benjamin W. and Hoehn, Jessica R. and Elby, Andrew and Finkelstein, Noah D. and Gupta, Ayush},
 doi = {10.1186/s40594-019-0187-y},
 issn = {2196-7822},
 journal = {International Journal of STEM Education},
 number = {1},
 pages = {31},
 title = {Splits in students’ beliefs about learning classical and quantum physics},
 volume = {6},
 year = {2019}
}

@article{ekert1991quantum,
 author = {Ekert, Artur K},
 journal = {Physical review letters},
 number = {6},
 pages = {661},
 publisher = {APS},
 title = {Quantum cryptography based on Bell’s theorem},
 volume = {67},
 year = {1991}
}

@article{borish_seeing_2023,
 author = {Borish, Victoria and Lewandowski, H. J.},
 doi = {10.1103/PhysRevPhysEducRes.19.020144},
 issn = {2469-9896},
 journal = {Physical Review Physics Education Research},
 number = {2},
 pages = {020144},
 title = {Seeing quantum effects in experiments},
 volume = {19},
 year = {2023}
}

@article{weisberg_embodied_2017,
 author = {Weisberg, Steven M. and Newcombe, Nora S.},
 doi = {10.1186/s41235-017-0071-6},
 issn = {2365-7464},
 journal = {Cognitive Research: Principles and Implications},
 number = {1},
 pages = {38, s41235--017--0071--6},
 shorttitle = {Embodied cognition and {STEM} learning},
 title = {Embodied cognition and {STEM} learning: overview of a topical collection in {CR}:{PI}},
 volume = {2},
 year = {2017}
}

@article{vlachopoulos_effect_2017,
 author = {Vlachopoulos, Dimitrios and Makri, Agoritsa},
 doi = {10.1186/s41239-017-0062-1},
 issn = {2365-9440},
 journal = {International Journal of Educational Technology in Higher Education},
 number = {1},
 pages = {22},
 shorttitle = {The effect of games and simulations on higher education},
 title = {The effect of games and simulations on higher education: a systematic literature review},
 volume = {14},
 year = {2017}
}

@book{aubussonMetaphorAnalogyScience2006a,
 author = {Aubusson, Peter and Harrison, Allan G. and Ritchie, Stephen M.},
 doi = {10.1007/1-4020-3830-5},
 isbn = {978-1-4020-3829-7, 978-1-4020-3830-3},
 langid = {english},
 location = {Dordrecht},
 number = {30},
 pagetotal = {210},
 publisher = {Springer Netherlands},
 series = {Science \& {{Technology Education Library}}},
 title = {Metaphor and {{Analogy}} in {{Science Education}}},
 year = {2006}
}

@article{buckleyModelBasedTeachingLearning2004,
 author = {Buckley, Barbara C. and Gobert, Janice D. and Kindfield, Ann C. H. and Horwitz, Paul and Tinker, Robert F. and Gerlits, Bobbi and Wilensky, Uri and Dede, Chris and Willett, John},
 doi = {10.1023/B:JOST.0000019636.06814.e3},
 issn = {1059-0145, 1573-1839},
 journaltitle = {Journal of Science Education and Technology},
 langid = {english},
 number = {1},
 pages = {23--41},
 shortjournal = {Journal of Science Education and Technology},
 shorttitle = {Model-{{Based Teaching}} and {{Learning}} with {{BioLogica}}™},
 title = {Model-{{Based Teaching}} and {{Learning}} with {{BioLogica}}™: {{What Do They Learn}}? {{How Do They Learn}}? {{How Do We Know}}?},
 volume = {13},
 year = {2004}
}

@article{gentnerStructuremappingTheoreticalFramework1983,
 author = {Gentner, D},
 doi = {10.1016/S0364-0213(83)80009-3},
 issn = {03640213},
 journaltitle = {Cognitive Science},
 langid = {english},
 number = {2},
 pages = {155--170},
 shortjournal = {Cognitive Science},
 shorttitle = {Structure-Mapping},
 title = {Structure-Mapping: {{A}} Theoretical Framework for Analogy},
 volume = {7},
 year = {1983}
}

@article{castro-alonsoResearchAvenuesSupporting2024,
 author = {Castro-Alonso, Juan C. and Ayres, Paul and Zhang, Shirong and De Koning, Björn B. and Paas, Fred},
 doi = {10.1007/s10648-024-09847-4},
 issn = {1040-726X, 1573-336X},
 journaltitle = {Educational Psychology Review},
 langid = {english},
 number = {1},
 pages = {10},
 shortjournal = {Educ Psychol Rev},
 title = {Research {{Avenues Supporting Embodied Cognition}} in {{Learning}} and {{Instruction}}},
 volume = {36},
 year = {2024}
}

@article{hutchinsMaterialAnchorsConceptual2005,
 author = {Hutchins, Edwin},
 doi = {10.1016/j.pragma.2004.06.008},
 issn = {03782166},
 journaltitle = {Journal of Pragmatics},
 langid = {english},
 number = {10},
 pages = {1555--1577},
 shortjournal = {Journal of Pragmatics},
 title = {Material Anchors for Conceptual Blends},
 volume = {37},
 year = {2005}
}

@article{kirshDistinguishingEpistemicPragmatic1994,
 author = {Kirsh, David and Maglio, Paul},
 doi = {10.1207/s15516709cog1804_1},
 issn = {0364-0213, 1551-6709},
 journaltitle = {Cognitive Science},
 langid = {english},
 number = {4},
 pages = {513--549},
 shortjournal = {Cognitive Science},
 title = {On {{Distinguishing Epistemic}} from {{Pragmatic Action}}},
 volume = {18},
 year = {1994}
}

@article{bennettQuantumCryptographyPublic2014,
 author = {Bennett, Charles H. and Brassard, Gilles},
 doi = {10.1016/j.tcs.2014.05.025},
 issn = {03043975},
 journaltitle = {Theoretical Computer Science},
 langid = {english},
 pages = {7--11},
 shortjournal = {Theoretical Computer Science},
 shorttitle = {Quantum Cryptography},
 title = {Quantum Cryptography: {{Public}} Key Distribution and Coin Tossing},
 volume = {560},
 year = {2014}
}

@incollection{harrison2006teaching,
  title={Teaching and learning with analogies: Friend or foe?},
  author={Harrison, Allan G and Treagust, David F},
  booktitle={Metaphor and analogy in science education},
  pages={11--24},
  year={2006},
  publisher={Springer}
}

\end{document}